# Field-induced magnetic phase transitions and memory effect in bilayer ruthenate $Ca_3Ru_2O_7$ with Fe substitution

M. Zhu[1], T. Hong[2], J. Peng[3], T. Zou[1], Z. Q. Mao[4], X. Ke[1*]

[1]*Department of Physics and Astronomy, Michigan State University, East Lansing, Michigan 48824, USA*

[2]*Neutron Scattering Division, Oak Ridge National Laboratory, Oak Ridge, Tennessee 37831, USA*

[3]*Collaborative Innovation Center of Advanced Microstructures, Laboratory of Solid State Microstructures, School of Physics, Nanjing University, Nanjing 210093, People's Republic of China*

[4]*Department of Physics and Engineering Physics, Tulane University, New Orleans, Louisiana 70118, USA*

* Corresponding author: Email address ke@pa.msu.edu

**Abstract**

Bilayer ruthenate $Ca_3(Ru_{1-x}Fe_x)_2O_7$ ($x = 0.05$) exhibits an incommensurate magnetic soliton lattice driven by the Dzyaloshinskii-Moriya interaction. Here we report complex field-induced magnetic phase transitions and memory effect in this system via single-crystal neutron diffraction and magnetotransport measurements. We observe first-order incommensurate-to-commensurate magnetic transitions upon applying the magnetic field both along and perpendicular to the propagation axis of the incommensurate spin structure. Furthermore, we find that the metastable states formed upon decreasing the magnetic field depend on the temperature and the applied field orientation. We suggest that the observed field-induced metastability may be ascribable to the quenched kinetics at low temperature.

## 1.    Introduction

Chiral helimagnets with long-period spatially modulated magnetic structures, stabilized by Dzyaloshinskii-Moriya (DM) interaction, have attracted revived interest in recent years.  On the one hand, it has been revealed that ferroelectricity can be driven by the noncollinear spin alignments, such as in rare-earth manganites $RMnO_3$ (R = Tb, Dy) and hexaferrites $(Ba,Sr)_2Zn_2Fe_{12}O_{22}$. Giant magnetoelectric coupling can be observed concomitant with the magnetic-field-induced incommensurate-to-commensurate spin structure transitions [1,2,3]. On the other hand, the field-induced magnetic structures of DM helimagnets can be very complex, and novel skyrmion phase may stabilize in a narrow regime in the field-temperature phase diagram [4,5]. Thus, incommensurate magnetic structures and the associated phase transitions deserve close investigations.

Recently it has been reported that incommensurate magnetic structures in a bilayer ruthenate $Ca_3Ru_2O_7$ can be induced by modest chemical substitution of other transition-metals for Ru [6]. The parent compound crystallizes in a non-centrosymmetric space group (No. 36 $Bb2_1m$) [7], and exhibits commensurate collinear antiferromagnetic structures below $T_N$ = 56 K. The easy axis of the staggered magnetic moments changes from the $a$ axis to the $b$ axis (denoted as AFM-a and AFM-b, respectively, as sketched in Fig. 1(a)) across a metal-insulator transition at $T_{MIT}$ = 48 K [8]. Upon 5% Fe substitution, the ground state of $Ca_3(Ru_{0.95}Fe_{0.05})_2O_7$ exhibits coexistence of the commensurate AFM-b type and an incommensurate magnetic soliton lattice below $T_{MIT}$ = 43 K (commensurate + incommensurate) [6]. The incommensurate magnetic component is a distorted cycloidal structure propagating along the $a$ axis, with a nearly temperature-independent propagation wave vector $\mathbf{q}_{ic}$ = ($\delta$ 0 1), $\delta \cong \pm0.017$, [Fig. 1(c)], as evidenced by the observation of the third-order harmonic magnetic Bragg peaks in neutron diffraction measurements [6]. Owing to the non-centrosymmetric crystal symmetry, DM interaction is expected to play an essential role in the formation of the incommensurate magnetic soliton lattice in Fe-doped $Ca_3Ru_2O_7$ [9]. The incommensurability $\delta$ of $Ca_3(Ru_{0.95}Fe_{0.05})_2O_7$ is rather small, indicating a long period of the incommensurate phase of ~58 unit cells or ~315 Å along the $a$ axis, which is typical for incommensurate magnetic structures stabilized by DM interactions.

The evolution of the magnetic soliton lattice phase in the presence of magnetic field is an intriguing problem to investigate, as it essentially reflects the interplay among exchange interactions, magnetic anisotropy, Zeeman interaction and DM interactions in this system. It is well-known that the magnetic phase transitions in DM helimagnets are rather complicated, depending on the orientation of the external magnetic field [10]. For instance, in the prototypical DM chiral helimagnet $Ba_2CuGe_2O_7$, a double-$k$ structure has been observed when the magnetic field is applied along the $c$ axis at $B_c = 2.1$ T, above which the system exhibits a commensurate antiferromagnetic structure [11]. In contrast, for an in-plane magnetic field, an incommensurate-to-commensurate magnetic transition occurs at $B_c = 9$ T and the double-$k$ structure is absent [12]. In FeGe, distinct modulated states have been observed in the small angle neutron scattering experiments with the magnetic field applied either transverse or longitudinal to the neutron beam [13]. Recently, neutron diffraction measurements on $Ca_3(Ru_{0.95}Fe_{0.05})_2O_7$ have revealed a first-order incommensurate-to-commensurate spin structure transition with the magnetic field applied along the $b$ axis, perpendicular to the spin propagation axis (i.e., $a$ axis) [14]. A metastable incommensurate magnetic soliton lattice with a much smaller incommensurability than that of the equilibrium state after zero field cooling has been discovered upon decreasing the magnetic field [14]. As in other DM helimagnets, complex magnetic phases may be induced with magnetic field applied along different crystallographic axes of $Ca_3(Ru_{0.95}Fe_{0.05})_2O_7$. In addition, whether or not the metastability is generic and how it depends on the orientation of applied magnetic field remain unresolved. And further studies are desirable to provide clues on the physical origin of the metastability phenomenon.

In this paper, we study field-induced spin structure transitions of $Ca_3(Ru_{1-x}Fe_x)_2O_7$ ($x = 0.05$) upon applying the magnetic field both parallel and perpendicular to the spin propagation direction ($a$ axis). A first-order incommensurate-to-commensurate magnetic transition is observed for $B$ // a, with much smaller critical field $B_c$ than that for $B$ // $b$. For $B$ // $a$, we find that the zero-field state after the magnetic field is swept down either maintains the high-field phase (CAFM-a), or transforms into an intermediate incommensurate magnetic structure with the incommensurability $\delta$ smaller than that of the equilibrium state ($\delta \cong \pm 0.017$), depending on the measurement temperature. The former is in sharp contrast to the metastability observed for $B$ // $b$ reported previously [14]. However, the similar characteristic temperatures associated with these metastable states for both applied field orientations suggest that these irreversible behaviors, though phenomenologically distinct, may have the same origin.

## 2.    Methods

The single crystals of $Ca_3(Ru_{1-x}Fe_x)_2O_7$ ($x = 0.05$) were synthesized by the floating zone method. Magnetotransport measurements were performed using the Physical Property Measurement System (PPMS, Quantum Design). Neutron diffraction experiments were carried out using the CG-4C cold neutron triple-axis spectrometer (CTAX) at the High Flux Isotope Reactor (HFIR) in Oak Ridge National Laboratory (ORNL). The energies of the incident and scattered neutrons were chosen as 5 meV to have good resolution in momentum transfer space. Contamination from higher-order reflections was removed by a cooled Be filter placed between the sample and the analyzer. The single-crystal sample was aligned in the horizontal ($H$ 0 $L$) scattering plane, where $H$ and $L$ are in reciprocal lattice units (r.l.u.) $2\pi/a = 1.17$ Å$^{-1}$ and $2\pi/c = 0.32$ Å$^{-1}$. The sample was mounted inside a cryomagnet with the horizontal magnetic field applied along the $a$ axis up to 4 T. Stray field has been investigated independently prior to the experiment and is not a concern within this field range. The magnet contains two pairs of windows for the incident and outgoing neutron beams. Constraint by the scattering geometry and the size of the window, only magnetic reflections around $\mathbf{q}_c = (0\ 0\ 1)$ are accessible in the neutron diffraction study. The intensity of neutrons is presented in the unit of counts per monitor count unit (mcu), where 1 mcu corresponds to ~1 sec.

## 3. Results

Figure 2(a) shows the normalized out-of-plane resistivity $\rho_c$ as a function of magnetic field at $T = 15$ K, as the magnetic field is applied along the $a$ and $b$ axis, respectively. The measurements were performed after the sample was initially cooled in zero magnetic field (ZFC) for each case. For $B\ //\ b$, the magnetoresistance displays a minimum at $B_c^\uparrow = 5 \pm 0.25$ T as the magnetic field increases, corresponding to the incommensurate-to-commensurate spin structure transition to CAFM-b revealed by neutron diffraction measurements [14]. The transition is of first order, as indicated by the appearance of hysteresis as the magnetic field increases and decreases ($B_c^\downarrow = 4.5 \pm 0.25$ T), in agreement with the neutron diffraction data [14]. Intriguingly, when the magnetic field is applied along the $a$ axis, a first-order transition is also observed at $B_c^\uparrow = 2.25 \pm 0.25$ T, in sharp contrast to that in the parent compound where no anomaly is observed in this field range [15,16]. This transition is likely to be associated with a field-induced spin structure transition as well. For comparison, Figure 2(b) shows the data taken at $T = 55$ K, where the material displays metallic state with an AFM-a type magnetic structure at zero field [6]. One can see that the features are completely different from that observed at 15 K. The magnetoresistance increases continuously as the magnetic field is along the *b* axis, whereas it exhibits a sudden jump for *B // a*. In the AFM-a phase, one would expected a first-order spin-flop or spin-flip transition when the magnetic field is along the easy axis (*a* axis), but a continuous transformation into the fully polarized state (PM) if the magnetic field is along the hard axis (*b* axis) [8]. The observations at 55 K seem to be well explained by this scenario.

Interestingly, the magnetoresistance data [Fig. 2(a)] for both *B // a* and *B // b* axes, which are essentially dominated by the spin scattering process of different magnetic structures, also exhibit irreversible behaviors. The zero-field resistance after applying the magnetic field up to 9 T (defined as $0{\downarrow}T$) is much greater than that measured after initially ZFC the sample from a temperature above $T_N$ (defined as $0{\uparrow}T$). The emergence of such hysteresis implies the formation of metastable magnetic phases when the magnetic field decreases, as discovered by the recent neutron diffraction study [14]. Similar to that induced with the magnetic field applied along the *b* axis, the metastable phase induced for *B // a* is also quite persistent. The change in the resistance is found to be less than ~0.1% in the time scale up to ~$10^4$ sec, and further cycling the magnetic field between 0 and 9 T gives rise to reversible phase transitions between the low-field metastable phases and the high-field ones. Such a magnetic memory effect can be erased only by heating the material up to high enough temperature, which suggests that thermal fluctuations play an essential role. Figure 3(a)-3(d) shows the irreversible field-induced phase transitions for *B // a* at representative temperatures. Because of the history effect, each measurement was done after heating the sample to a temperature above $T_N$ then ZFC to the measurement temperature to clean up the magnetic memory. Below 26 K, as the magnetic field decreases from 9 T, the magnetoresistance monotonically increases and becomes saturated, giving rise to a large hysteresis. On the contrary, for *T* > 26 K, the magnetoresistance data [Fig. 3(b)-3(d)] shows a decrease at the critical field $B_c^{\downarrow}$ (marked by green arrows) as the magnetic field decreases, suggesting that the phase transformation starts to occur. In addition, the remnant magnetoresistance at zero field ($0{\downarrow}T$) decreases with increasing temperature, and finally disappears at $T \cong 34$ K [Fig. 3(d)], which may suggest that the equilibrium phase (commensurate + incommensurate, $\delta \cong \pm 0.017$) has been recovered.

In order to determine the magnetic structure of the field-induced phase below $T_{MIT}$ and the metastable phase at $B < B_c^{\downarrow}$, we have carried out neutron diffraction measurements with the magnetic field applied along the $a$ axis. Due to the limited access to the reciprocal space with the horizontal-field cryomagnet, we only focus on the magnetic reflections around $\mathbf{q}_c = (0\ 0\ 1)$. Figure 4(a) and 4(b) show the contour maps of the neutron diffraction intensity of scans across $\mathbf{q}_c = (0\ 0\ 1)$ along the [1 0 0] direction at $T = 15$ K (every 0.5 T from 0 to 4 T). At zero field, the system exhibits coexistence of the commensurate magnetic peaks $\mathbf{q}_c = (0\ 0\ 1)$ and incommensurate ones $\mathbf{q}_{ic} = (\delta\ 0\ 1)$, $\delta = \pm 0.017$, in agreement with the previous study [6]. As the magnetic field increases, the incommensurate peaks are suppressed at a critical field $B_c^{\uparrow} = 2.25 \pm 0.25$ T (defined as the point where the incommensurate peaks are gone completely), much smaller than that when $B\ /\!/\ b$, and the incommensurability $\delta$ keeps nearly constant during the phase transition. In the meantime, the intensity of the commensurate one is significantly enhanced and the peak width becomes broader. As the magnetic field increases further, the (0 0 1) peak intensity gets stronger and the peak width of (0 0 1) gets narrower. However, the integrated intensity decreases slightly, implying that the antiferromagnetic staggered magnetization is gradually suppressed by the magnetic field. Note that the reflection condition of the crystal symmetry of $Ca_3(Ru_{0.95}Fe_{0.05})_2O_7$ ($Bb2_1m$, No. 36) requires that both $H$ and $L$ are even, or the sum $H+L$ is even [7]. The emergence of reflections at (0 0 1) suggests that the system is in an antiferromagnetic phase, similar to the field-induced states in the pristine and other doped $Ca_3Ru_2O_7$ [8,17,18]. Although only one magnetic reflection is available in this experiment, considering that in all other related studies on $Ca_3Ru_2O_7$ this magnetic reflection corresponds to an AFM-a or AFM-b type structure [8,14,17], we propose that the field-induced magnetic structure is CAFM-a, a superposition of AFM-b and a ferromagnetic component along the *a* axis, as shown in Fig. 1(b).

The irreversibility of the field-induced magnetic transition is clearly observed in the neutron diffraction measurements. As the magnetic field decreases, surprisingly, both the peak width and peak intensity of the commensurate (0 0 1) magnetic reflection stay nearly unchanged down to 0 T, and the incommensurate ones do not reemerge at $T = 15$ K. This agrees well with the magnetoresistance measurements, where the critical field $B_c^{\downarrow}$, i.e., the phase transformation into the equilibrium commensurate + incommensurate phase, is absent [Fig. 2(a)]. This observation is different from the previous report for $B \mathbin{/\mkern-5mu/} b$, where the system transforms into a metastable state with much smaller incommensurability $\delta \cong \pm 0.004$ ($T = 15$ K) [14]. Figure 4(c) and 4(d) show the scans at $B = 0\uparrow$, $4\downarrow$ and $0\downarrow$T. Inset show the scans at $B = 0\uparrow$ T and $0\downarrow$T as the magnetic field is applied along the *b* axis for comparison. All the incommensurate and commensurate magnetic peaks can be well fitted by Gaussian functions (a broad peak is added in Fig. 4(c) to account for the broad diffuse scattering intensity, as discussed in detail in Ref. [14]). The peak widths of all the peaks are not resolution limited (the resolution is denoted by the short gray line). Particularly the width of (0 0 1) at $B = 4$ T is much broader than that in the $0\uparrow$T phase, which suggests a much shorter correlation length ~210a (a is the lattice constant) in the field-induced CAFM-a phase and the metastable state at $B = 0\downarrow$ T.

We also performed neutron diffraction measurements at a higher temperature $T = 34$ K, where a critical field $B_c^{\downarrow}$ is seen in the magnetoresistance measurements [Fig. 3(b)-3(d)], as shown in Fig. 5(a) and 5(b). Similarly, the incommensurate-to-commensurate magnetic transition occurs at $B_c^{\uparrow} =$ $2.75 \pm 0.25$ T, close to that shown in the magnetoresistance measurements [Fig. 3(d)]. However, as the magnetic field decreases, at $B_c^\downarrow = 2$ T, the commensurate (0 0 1) peak becomes weaker and two incommensurate peaks appear at $\mathbf{q}_{ic} = (\delta\ 0\ 1)$, $\delta \cong \pm0.01$. Upon further reducing the magnetic field, the incommensurability $\delta$ gradually evolves toward that of the equilibrium state, which indicates a decrease in the period of the incommensurate magnetic structure. Nevertheless, the equilibrium value of the incommensurability $\delta \cong \pm0.017$ cannot be reached down to 0 T at this temperature. A complete recovery of the initial incommensurability occurs in the measurements done at $T_g = 37$ K, the same as that when the magnetic field is along the $b$ axis [14]. The small discrepancy in $T_g$ where the equilibrium state is recovered determined by neutron diffraction and magnetotransport measurements discussed previously [Fig. 3(d)] might be due to the fact that the incommensurability can be better revolved by neutron diffraction. Figure 5(c) and 5(d) shows the representative scans at $B = 0\uparrow, 4\uparrow$ T and $B = 1.5\downarrow, 0\downarrow$ T, respectively. The peak width of (0 0 1) at $B > B_c^\downarrow$ is not resolution limited, corresponding to a correlation length of ~450a (a is the lattice constant). In the metastable state, the correlation length of the commensurate phase is comparable to that of the CAFM-a phase, but the incommensurate peaks are broader and become slightly narrower as the magnetic field decreases.

To make a further comparison with the $B // b$ case, we also studied the evolution of the $0\downarrow T$ metastable phase as a function of temperature. The metastable state was obtained using the same procedure as the $0\downarrow T$ phase discussed in Fig. 4(b). Namely, we first ZFC the sample to 15 K and applied 4 T magnetic field along the $a$ axis. The magnetic field was then reduced to zero. Figure 6(a) shows the contour map of the neutron intensity obtained by scanning across the $\mathbf{q}_c = (0\ 0\ 1)$ peak along the [1 0 0] direction at various temperatures (every 2 K from 15 to 50 K). Upon warming the magnetic reflection (0 0 1) almost doesn't change until $T' = 31$ K, where the incommensurate peaks start to appear and move towards the equilibrium state with $\mathbf{q}_{ic} = (\delta\ 0\ 1)$, $\delta \cong \pm 0.017$. As soon as the incommensurate peaks show up, the intensity of the commensurate (0 0 1) peak becomes much weaker, indicating that the commensurate magnetic structure (CAFM-a) transforms into the incommensurate ones. Fig. 6(b) shows the scans at selected temperatures $T = 15$, 34 and 37 K. One clearly see that the equilibrium states is reached at $T_g = 37$ K, similar to the previous study on $B \mathbin{/\mkern-3mu/} b$ [14]. In addition, the peak width of commensurate (0 0 1) at $T = 34$ K is much narrower than that at 15 K, which implies that the correlation length becomes much longer (~440a with a $\cong 5.37$ Å being the lattice constant). In contrast, the peak width of the incommensurate peaks is broader and corresponds to a correlation of ~230a. For $T > 37$ K, the behavior of the magnetic Bragg peaks is the same as that of the equilibrium 0↑T phase [6].

## 4. Discussions

In $Ca_3(Ru_{1-x}Fe_x)_2O_7$ ($x = 0.05$), the observation of first-order magnetic-field-induced spin structure transitions as the magnetic field is applied both along the $a$ and $b$ axis is characteristic of the complex incommensurate magnetic soliton lattice phase induced upon Fe. The feature is in sharp contrast to that in the parent compound [15,16]. In the pure $Ca_3Ru_2O_7$, below $T_{MIT}$, the material displays collinear AFM-b type magnetic structure, with the staggered magnetization along the $b$ axis [8]. The magnetic field leads to a first-order spin-flop transition at $B_c = 6$ T ($T = 4$ K) only when the magnetic field is applied along the $b$ axis [16]. On the contrary, as the magnetic field is applied along the $a$ axis, the system continuously evolves towards the fully spin polarized state (PM) and no first-order spin structure transition is observed [8]. Note that although a field-induced phase transition is seen at 15 T, it is attributed to a change in the orbital occupancy [16]. As a DM helimagnet, the stabilization of the incommensurate magnetic soliton lattice in $Ca_3(Ru_{0.95}Fe_{0.05})_2O_7$ and its evolution as a function of temperature and magnetic field can be governed by the competition among Zeeman interaction, exchange interaction, DM interaction, and magnetocrystalline anisotropy, a mechanism that accounts for different types of spin structure transitions in various other DM magnets, such as $Ba_2CuGeO_7$ [19,12] and FeGe [13,20,21], etc. It is worth noting that the field-induced magnetic transitions for $B // b$ and $B // a$, though both have been identified as incommensurate-to-commensurate spin structure transitions by neutron diffraction measurements, are different magnetic phase transitions. The critical field for $B // a$ is rather small, compared with that of $B // b$ case, and the field-induced commensurate phases are distinct. However, a more detailed theory is required to interpret these observations quantitatively.

The observation of irreversibility in the field-induced incommensurate-to-commensurate magnetic transitions for both $B // a$ and $B // b$ implies its universality in the field-induced magnetic transitions in Fe-doped $Ca_3Ru_2O_7$. However, a closer comparison between the cases of $B // b$ and $B // a$, and with other magnetic systems has provided important implications on these metastable problems. It is worth noting that in the same material system $Ca_3(Ru_{0.95}Fe_{0.05})_2O_7$ the irreversible behaviors for $B // b$ and $B // a$ could be different. In the previous study where the magnetic field is applied along the $b$ axis, at $T = 15$ K the phase transformation into an incommensurate phase occurs as the magnetic field decreases, but gives rise to a much smaller incommensurability ($\delta = \pm 0.004$) than that of the equilibrium state ($\delta = \pm 0.017$) in the $0\downarrow T$ phase, which indicates that the soliton density in the magnetic soliton lattice phase cannot reach the equilibrium value. And it has been proposed that the pinning of the domain walls by the magnetic Fe impurities may be responsible for this metastability [14]. In contrast, at $T = 15$ K for $B // a$, as shown in Fig. 4, the system maintains the high-field CAFM-a state down to zero field ($0\downarrow$ T), suggesting that the phase transformation even cannot occur. This makes it a little puzzling as the magnetic impurities are expected to enhance the inhomogeneous nucleation process, which should have facilitated the formation of the new phase. Interestingly, these two different irreversible behaviors have been reported in a number of separate magnetic systems, such as $TbMnO_3$ [22], $DyMn_2O_5$ [23], UNiAl [24], where different incommensurability or even an alternative modulation wave vector emerges in the $0\downarrow T$ phase, and Cr [25], $Ba_{0.5}Sr_{1.5}Zn_2(Fe_{1-x}Al_x)_{12}O_{22}$ [26] and $NaFe(WO_4)_2$ [27], where the high-pressure or high-field states are maintained in the $0\downarrow T$ state upon releasing the pressure or removing the magnetic field.

Despite of distinct metastable $0\downarrow T$ states discussed above, we speculate that these irreversible phenomena might be of the same origin. First, the irreversible behaviors become similar at $T = 34$ K, though being completely different at $T = 15$ K, i.e., the phase transformation into an incommensurate phase occurs as the magnetic field decreases for both $B$ // b and $B$ // a, and gives rise to a smaller incommensurability $\delta$ than that of the equilibrium state ($\delta = \pm0.017$) in the $0\downarrow T$ phase. Second, the characteristic temperatures associated with the metastability in $Ca_3(Ru_{0.95}Fe_{0.05})_2O_7$ are the same for both $B$ // a and $B$ // b. For instance, the critical temperature $T_g$ above which the system completely restores to the equilibrium state is $T_g = 37$ K for both cases. And the temperature when the metastable state at $B = 0\downarrow$ T starts to evolve toward the equilibrium one, $T' = 31$ K as shown in Fig. 6(a), is very similar to that with the magnetic field applied along $b$ axis [Fig. 4(a) in Ref. [14]].

A plausible qualitative explanation for the field-induced irreversibility in the magnetic modulation vector may be given from the perspective of the free energy landscape as a function of thermodynamic variables. Upon decreasing the magnetic field, due to the first-order nature of the phase transition, the nucleation of a new phase requires activation energy in order to overcome the intervening energy barriers between the local minima. However, thermal fluctuations at low temperature are not strong enough, which prevents the formation of large enough nucleus of the stable phase [28]. The critical field $B_c^\downarrow$ signifies the magnetic field at which the strength of the thermal energy $\sim k_B T$ and the height of the energy barrier U between the CAFM phase and the metastable phase become comparable. Such a mechanism was applied to account for the irreversibility observed in phase-separated manganites $Nd_{0.5}Sr_{0.5}MnO_3$ and $Pr_{1-x}Ca_xMnO_3$ where large hysteresis is observed in the phase transition between a ferromagnetic metallic state and an antiferromagnetic charge-ordered phase driven by a magnetic field [29,30]. Following the analysis in Ref. [29], the temperature dependence of $B_c^\downarrow$ obtained by magnetoresistance measurements, as summarized in the inset of Fig. 3(a), can be approximately scaled using an empirical function $T \propto \left| B - B_c^{(0)} \right|^\beta$, where $B_c^{(0)}$ is the critical field when U becomes zero [29]. The fact that $B_c^\downarrow$ becomes greater as the temperature increases suggests that the height of the energy barrier U decreases as the magnetic field is swept down, in analogy to that in the manganite systems [29]. The value of the critical exponent β is sensitively dependent on the choice of $B_c^{(0)}$. In the inset of Fig. 3(a), the solid line represents the fit with $\beta = 1.5$.

## 5. Conclusions

In summary, we have investigated the magnetic-field-induced phase transitions of $Ca_3(Ru_{1-x}Fe_x)_2O_7$ ($x = 0.05$) with the magnetic field applied both along and perpendicular to the propagation direction of the incommensurate spin structure. First-order incommensurate-to-commensurate magnetic transitions into canted antiferromagnetic structures, either CAFM-a for $B \mathbin{/\mkern-5mu/} a$ or CAFM-b for $B \mathbin{/\mkern-5mu/} b$, have been observed. The field-induced transitions display distinct irreversible behaviors below a characteristic temperature $T_g$ and gives rise to persistent metastable states, where the system either remains in the high-field phase or transforms into an incommensurate state but with smaller incommensurability. These observations can be qualitatively described in the framework of frozen kinetics at low temperature, which prevents the system from overcoming the energy barriers and reaching the equilibrium state.

**Acknowledgement**

The work at Michigan State University was supported by the National Science Foundation under Award No. DMR-1608752 and the start-up funds from Michigan State University. The work at Tulane is supported by the U.S. Department of Energy under EPSCoR Grant No. DE-SC0012432 with additional support from the Louisiana Board of Regents. A portion of this research used resources at the High Flux Isotope Reactor, a DOE Office of Science User Facility operated by the Oak Ridge National Laboratory. Work at Nanjing University was supported by the National Natural Science Foundation of China (Grant No. 11304149).

**Figure captions**

Figure 1. Schematic diagrams of the magnetic structures in $Ca_3(Ru_{1-x}Fe_x)_2O_7$ ($x = 0.05$). (a) AFM-a and AFM-b: collinear antiferromagnetic structures with the magnetic moments along the $a$ and $b$ axis, respectively. (b) CAFM-b: canted antiferromagnetic structure consisting of an AFM-a component and a ferromagnetic one along the $b$ axis. CAFM-a: canted antiferromagnetic structure consisting of an AFM-b component and a ferromagnetic one along the $a$ axis. (c) In-plane view of one of the bilayers of the incommensurate (ICM) magnetic soliton lattice. The dashed square in magenta represents one unit cell. The orange rectangle with rounded corners encloses the domain wall. The spin rotation of the magnetic domain wall is proposed to be 180°, owing to a magnetic anisotropy of order 2. The blue and orange arrows stand for the direction of the magnetic field $B$ // a and $B$ // b, respectively.

Figure 2. (a) Normalized out-of-plane resistivity $\rho_c$ of $Ca_3(Ru_{1-x}Fe_x)_2O_7$ ($x = 0.05$) as a function of the magnetic field at (a) $T = 15$ K and (b) $T = 55$ K. The magnetic field is applied along the $a$ and $b$ axis, respectively.

Figure 3. Normalized out-of-plane resistivity $\rho_c$ of $Ca_3(Ru_{1-x}Fe_x)_2O_7$ ($x = 0.05$) as a function of the magnetic field at representative temperatures. The upper and lower critical fields are denoted by red and green arrows, respectively. Inset shows the lower critical field $B_c^1$ as a function of temperature. The red solid line represents a fit using an empirical function described in the text. The magnetic field is along the $a$ axis.

Figure 4. Contour maps of the neutron intensity of scans across the magnetic reflection $\mathbf{q}_c$ = (0 0 1) along the [1 0 0] direction at $T$ = 15 K as (a) the magnetic field increases from 0 to 4 T. (b) the magnetic field decreases from 4 to 0 T. (c),(d) Cuts of the scans taken at $B$ = 0↑, 4↓ and 0↓ T defined in the text. The solid and dashed lines are fitted curves using Gaussian functions. The gray line denotes the instrument resolution. The magnetic field is applied along the $a$ axis. Inset shows the scans at $B$ = 0↑ (blue) and 0↓ T (red) when magnetic field is along the $b$ axis at $T$ = 15 K.

Figure 5. Contour maps of the neutron intensity of scans across the magnetic reflection $\mathbf{q}_c$ = (0 0 1) along the [1 0 0] direction at $T$ = 34 K as (a) the magnetic field increases from 0 to 4 T. (b) the magnetic field decreases from 4 to 0 T. (c),(d) Cuts of the scans taken at $B$ = 0↑, 4↑ T and 1.5↓, 0↓ T defined in the text. The solid lines are fitted curves using Gaussian functions. The gray line denotes the instrument resolution. The magnetic field is applied along the $a$ axis.

Figure 6. (a) Contour map of the neutron intensity of scans across the magnetic reflection $\mathbf{q}_c$ = (0 0 1) along the [1 0 0] direction in the 0↓$T$ phase at various temperatures. (b) Cuts of the scans taken at T = 15, 34, and 37 K in the 0↓$T$ phase. The solid lines are fitted curves using Gaussian functions. The gray line represents the instrument resolution.

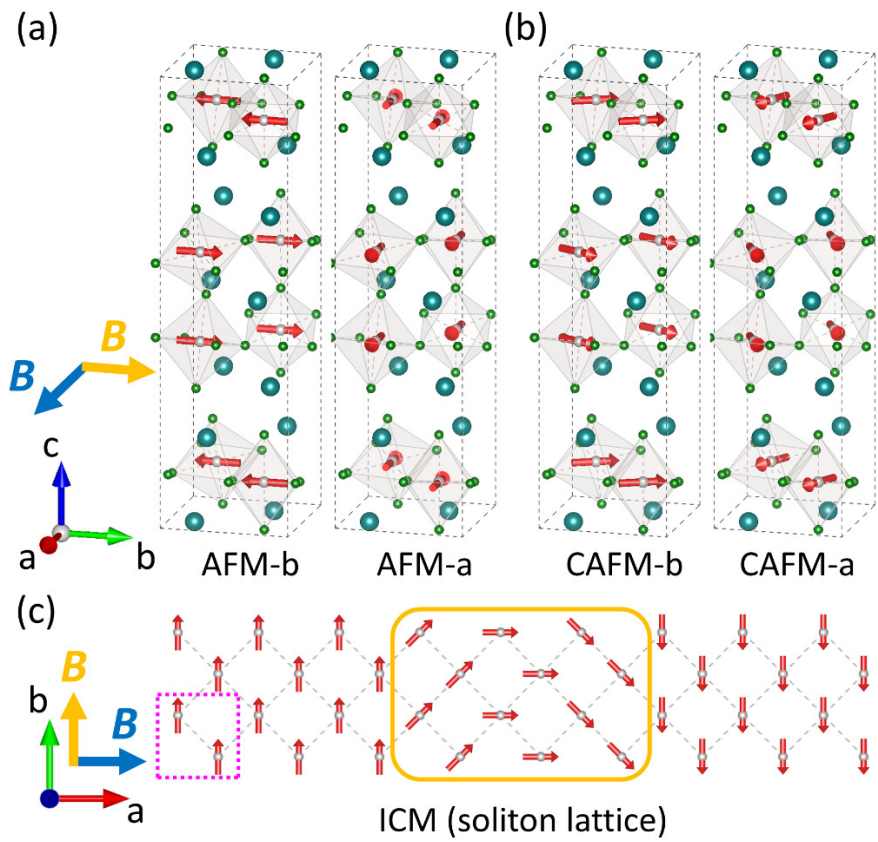

(a) AFM-b  AFM-a  (b) CAFM-b  CAFM-a

(c) ICM (soliton lattice)

**Figure 2**

M. Zhu et al.

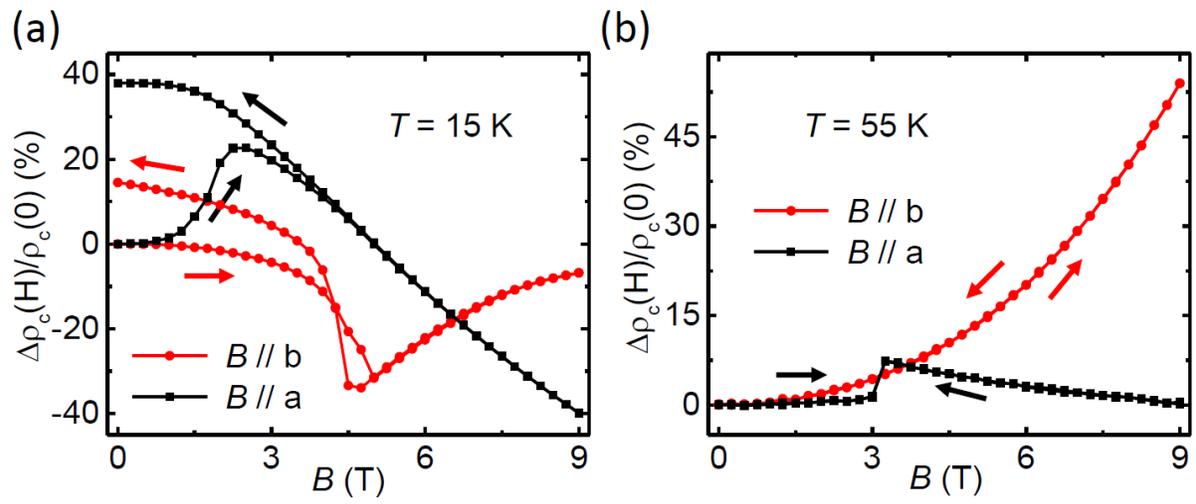

**Figure 3**

M. Zhu et al.

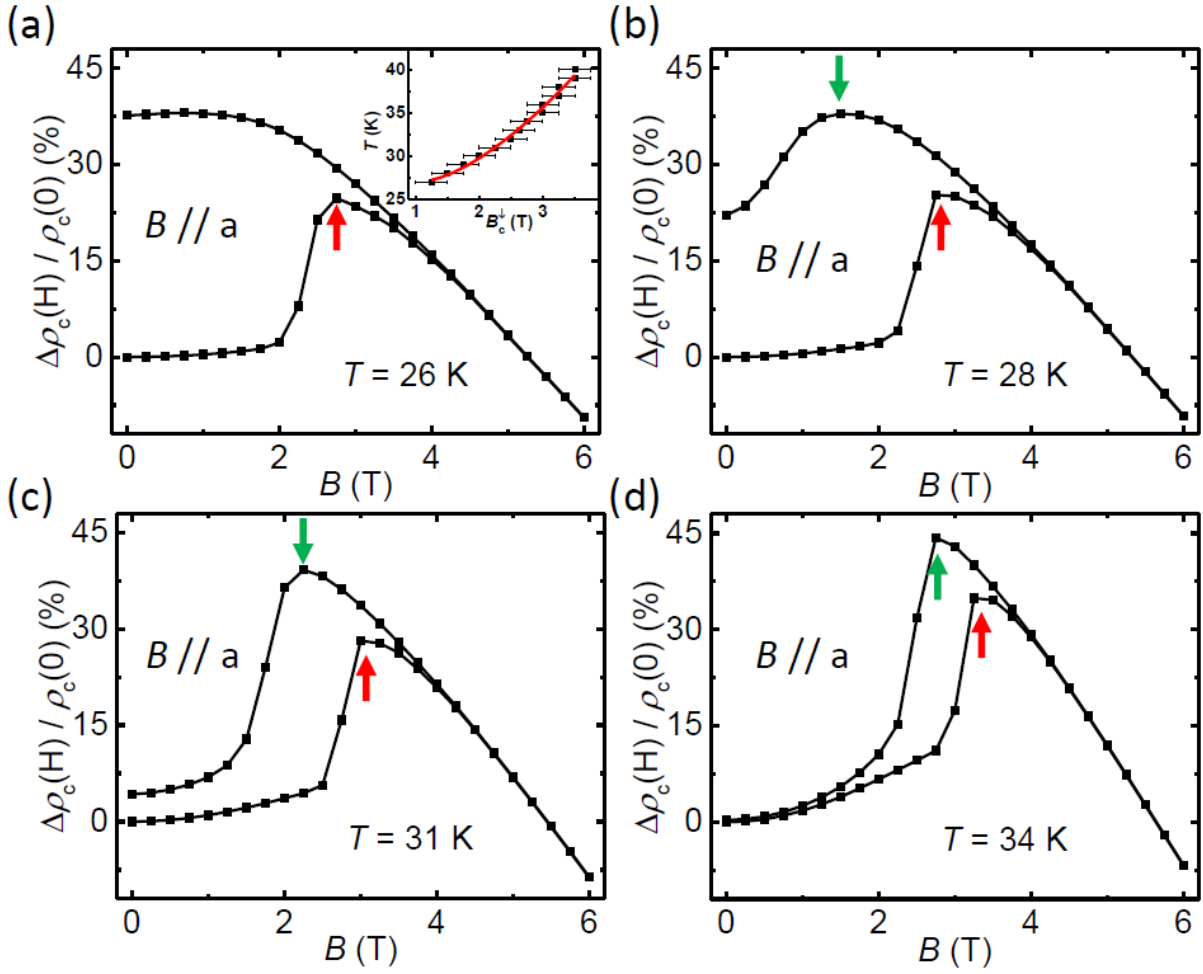

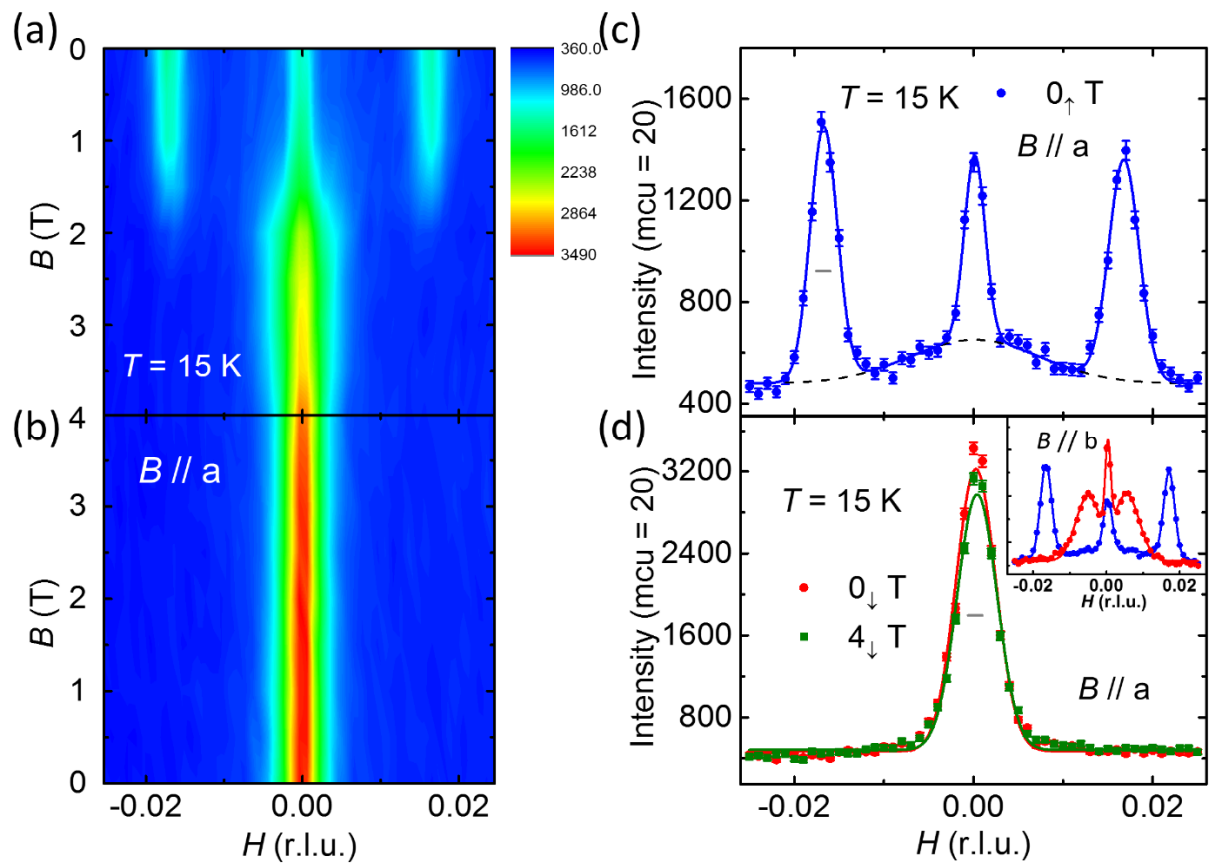

**Figure 5**

M. Zhu et al.

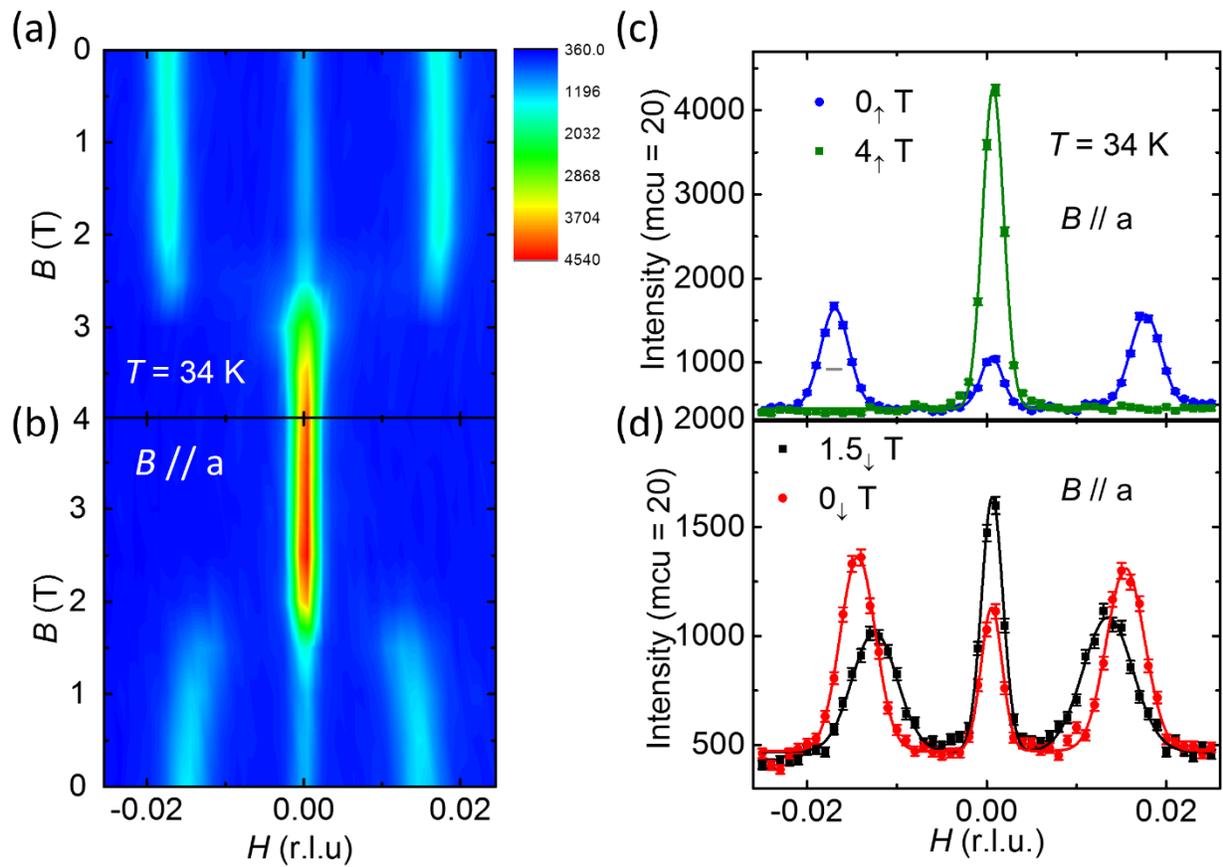

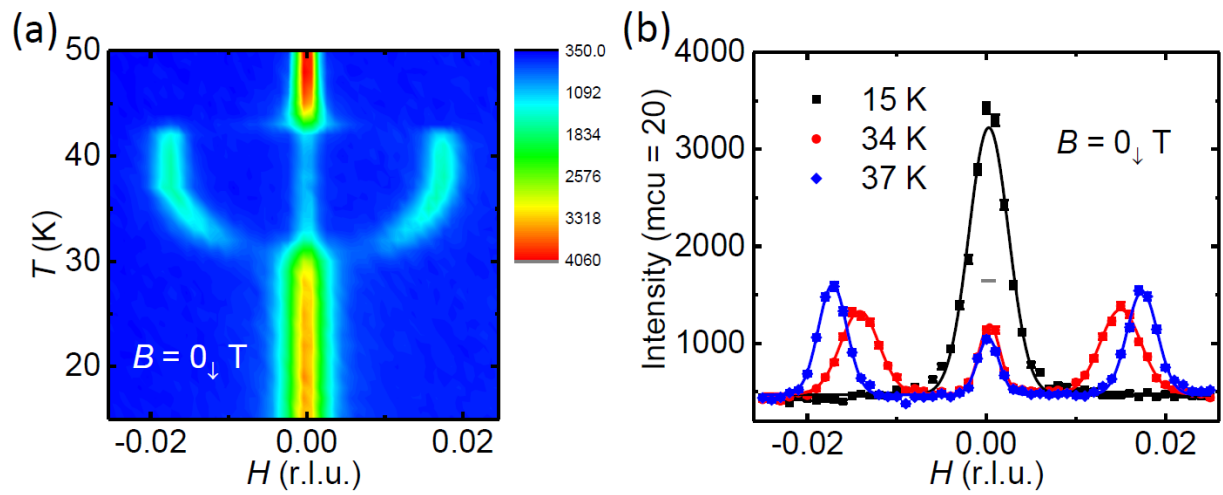